\documentclass[letterpaper,twocolumn,10pt]{article}
\usepackage{usenix-2020-09}

\usepackage{tikz}
\usepackage{amsmath}

\usepackage{filecontents}

\usepackage{tabularx}
\usepackage{booktabs}
\usepackage{tikz}
\usepackage{xcolor}
\usepackage{xspace}
\usepackage{tikz}
\usepackage{pifont}
\usepackage[english]{babel}
\usepackage{blindtext}
\usepackage{lipsum}
\usepackage{makecell}
\usepackage[ruled,vlined,linesnumbered]{algorithm2e}
\usepackage{booktabs}

\usepackage{filecontents}
\usepackage{xspace}
\usepackage{subfig}
\usepackage{xcolor}
\usepackage{hhline}
\usepackage{multirow}
\usepackage{soul}

\usepackage{array}
\usepackage{comment}
\usepackage{listings}
\usepackage{dsfont}
\usepackage{algorithmic}

\usepackage{xurl}
\usepackage{setspace}
\usepackage[flushleft]{threeparttable}
\usepackage{caption}
\usepackage{ltablex}

\usepackage{ifluatex}

\usepackage{outlines}

\usepackage{multirow}

\usepackage{hyperref}
\usepackage{hyperxmp}

\newtheorem{insight}{Insight}

\begin{document}

\newcommand*{\affmark}[1][*]{\textsuperscript{#1}}
\newcommand*{\affaddr}[1]{#1}

\title{{\name: Joint KV-Cache Compression and Eviction for Efficient LLM Serving}}

\author{
  \normalsize
  Shaoting Feng$^{*,1}$ \quad
  Yuhan Liu$^{*,1}$ \quad
  Hanchen Li$^{2}$ \quad
  Xiaokun Chen$^{3}$ \quad
  Samuel Shen$^3$ \quad
  Kuntai Du$^3$ \quad
  Zhuohan Gu$^4$ \\
  \normalsize
  Rui Zhang$^5$ \quad \hspace{-.15in}
  Yuyang Huang$^1$ \quad \hspace{-.15in}
  Yihua Cheng$^3$ \quad \hspace{-.15in}
  Jiayi Yao$^1$ \quad \hspace{-.15in}
  Qizheng Zhang$^6$ \quad \hspace{-.15in}
  Ganesh Ananthanarayanan$^7$ \quad \hspace{-.15in}
  Junchen Jiang$^{1,3}$ \\
  \normalsize
  $^1$University of Chicago \hspace{.1in} $^2$UC Berkeley \hspace{.1in} $^3$Tensormesh, Inc. \hspace{.1in} $^4$MIT \hspace{.1in} $^5$UC Santa Cruz \hspace{.1in} $^6$Stanford \hspace{.1in} $^7$Microsoft 
}


\newcommand{\edit}[1]{{\color{black} #1}}
\newcommand{\shan}[1]{{\color{red}(Shan: #1)}}
\newcommand{\shanedit}[1]{{\color{red} #1}} 
\newcommand{\hank}[1]{{\color{blue}(Hank: #1)}}
\newcommand{\mm}[1]{{\color{violet}(Michael:#1)}}
\newcommand{\jc}[1]{{\footnotesize\color{orange}{(JC: #1)}}}
\newcommand{\jcedit}[1]{{\color{orange} #1}} 
\newcommand{\yh}[1]{{\footnotesize\color{deepgreen}{(Yuhan: #1)}}}
\newcommand{\jiayi}[1]{{\color{brown}{(Jiayi: #1)}}}
\newcommand{\kt}[1]{{\color{violet}{(Kuntai: #1)}}}
\newcommand{\zgu}[1]{{\color{teal}{(Zhuohan: #1)}}}
\newcommand{\shaoting}[1]{{\color{violet}{(Shaoting: #1)}}}
\newcommand{\slshen}[1]{{\color{yellow}{(Samuel: #1)}}}
\newcommand{\todo}[1]{{\color{red}{(TODO: #1)}}}
\newcommand{\hcedit}[1]{{\color{black} #1}}
\definecolor{darkkhaki}{rgb}{0.74, 0.72, 0.42}
\newcommand{\hc}[1]{{\color{darkkhaki}{(LHC: #1)}}}
\definecolor{hhcolor}{RGB}{240, 35, 240}
\newcommand{\hh}[1]{{\color{hhcolor}{(Yihua: #1)}}}
\newcommand{\sr}[1]{{\color{cyan!70!blue}{(Siddhant: #1)}}}
\newcommand{\qz}[1]{{\color{purple}{(Qizheng: #1)}}}

\newcommand{\KV}{\ensuremath{KV}\xspace}
\newcommand{\Attention}{\ensuremath{A}\xspace}
\newcommand{\ForwardAttention}{\ensuremath{FA}\xspace}
\newcommand{\LayerIndex}{\ensuremath{i}\xspace}
\newcommand{\TokenIndex}{\ensuremath{j}\xspace}
\newcommand{\UserTokenIndex}{\ensuremath{j'}\xspace}
\newcommand{\ChunkIndex}{\ensuremath{n}\xspace}
\newcommand{\ChunkNum}{\ensuremath{N}\xspace}
\newcommand{\Func}{\ensuremath{F}\xspace}
\newcommand{\Pre}{\textrm{pre}\xspace}
\newcommand{\New}{\textrm{new}\xspace}
\newcommand{\Full}{\textrm{full}\xspace}
\newcommand{\CA}{\ensuremath{C}\xspace}

\newcommand{\KVD}{\ensuremath{\Delta_{\textrm{kv}}}\xspace}
\newcommand{\CAD}{\ensuremath{\Delta_{\textrm{attn}}}\xspace}

\newcommand{\ACA}{{ACA}\xspace}
\newcommand{\HCA}{{HKVD}\xspace}

\newcommand{\name}{\textsc{EvicPress}\xspace}
\newcommand{\promptcache}{PromptCache\xspace}
\newcommand{\HCF}{HCF tokens\xspace}

\newcommand{\vspacesize}{0.2cm}

\newcommand{\fillme}{{\bf XXX}\xspace}

\newcommand*\circled[1]{\tikz[baseline=(char.base)]{
            \node[shape=circle,fill,inner sep=2pt] (char) {\textcolor{white}{\footnotesize{#1}}};}}

\newcounter{packednmbr}
\newenvironment{packedenumerate}{\begin{list}{\thepackednmbr.}{\usecounter{packednmbr}\setlength{\itemsep}{0.5pt}\addtolength{\labelwidth}{-4pt}\setlength{\leftmargin}{2ex}\setlength{\listparindent}{\parindent}\setlength{\parsep}{1pt}\setlength{\topsep}{0pt}}}{\end{list}}
\newenvironment{packeditemize}{\begin{list}{$\bullet$}{\setlength{\itemsep}{0.5pt}\addtolength{\labelwidth}{-4pt}\setlength{\leftmargin}{2ex}\setlength{\listparindent}{\parindent}\setlength{\parsep}{1pt}\setlength{\topsep}{2pt}}}{\end{list}}
\newenvironment{packedpackeditemize}{\begin{list}{$\bullet$}{\setlength{\itemsep}{0.5pt}\addtolength{\labelwidth}{-4pt}\setlength{\leftmargin}{\labelwidth}\setlength{\listparindent}{\parindent}\setlength{\parsep}{1pt}\setlength{\topsep}{0pt}}}{\end{list}}
\newenvironment{packedtrivlist}{\begin{list}{\setlength{\itemsep}{0.2pt}\addtolength{\labelwidth}{-4pt}\setlength{\leftmargin}{\labelwidth}\setlength{\listparindent}{\parindent}\setlength{\parsep}{1pt}\setlength{\topsep}{0pt}}}{\end{list}}
\let\enumerate\packedenumerate
\let\endenumerate\endpackedenumerate
\let\itemize\packeditemize
\let\enditemize\endpackeditemize

\newcommand{\tightcaption}[1]{\vspace{-0.2cm}\caption{{\normalfont{\textit{{#1}}}}}\vspace{-0.2cm}}
\newcommand{\tightsection}[1]{\vspace{-0.3cm}\section{#1}\vspace{-0.2cm}}
\newcommand{\tightsectionstar}[1]{\vspace{-0.17cm}\section*{#1}\vspace{-0.08cm}}
\newcommand{\tightsubsection}[1]{\vspace{-0.25cm}\subsection{#1}\vspace{-0.1cm}}
\newcommand{\tightsubsubsection}[1]{\vspace{-0.01in}\subsubsection{#1}\vspace{-0.01cm}}

\newcommand{\eg}{{\it e.g.,}\xspace}
\newcommand{\ie}{{\it i.e.,}\xspace}
\newcommand{\etal}{{\it et.~al}\xspace}
\newcommand{\bigO}{\mathrm{O}}
\newcommand{\twlog}{w.l.o.g.\xspac}

\newcommand{\myparashort}[1]{\vspace{0.05cm}\noindent{\bf {#1}}~}
\newcommand{\mypara}[1]{\vspace{0.05cm}\noindent{\bf {#1}:}~}
\newcommand{\myparatight}[1]{\vspace{0.02cm}\noindent{\bf {#1}:}~}
\newcommand{\myparaq}[1]{\smallskip\noindent{\bf {#1}?}~}
\newcommand{\myparaittight}[1]{\smallskip\noindent{\emph {#1}:}~}
\newcommand{\question}[1]{\smallskip\noindent{\emph{Q:~#1}}\smallskip}
\newcommand{\myparaqtight}[1]{\smallskip\noindent{\bf {#1}}~}

\newcommand{\cmark}{\ding{51}}%
\newcommand{\xmark}{\ding{55}}%



\definecolor{backcolour}{rgb}{0.96,0.96,0.96}
\definecolor{codegray}{rgb}{0.5,0.5,0.5}
\definecolor{deepblue}{rgb}{0,0,0.6}
\definecolor{deepred}{rgb}{0.6,0,0}
\definecolor{deepgreen}{rgb}{0,0.5,0}
\lstdefinestyle{mystyle}{
    backgroundcolor=\color{backcolour},   
    commentstyle=\color{codegreen},
    morekeywords={self, True},
    keywordstyle=\color{deepblue},
    numberstyle=\tiny\color{codegray},
    emph={MyClass,__init__,EncodingType,Image},
    emphstyle=\color{deepred},
    stringstyle=\color{deepgreen},
    basicstyle=\ttfamily\footnotesize,
    breakatwhitespace=false,         
    breaklines=true,                 
    captionpos=b,                    
    keepspaces=true,                 
    numbers=left,                    
    numbersep=5pt,                  
    showspaces=false,                
    showstringspaces=false,
    showtabs=false,                  
    tabsize=1
}

\maketitle
\renewcommand{\thefootnote}{\fnsymbol{footnote}}
\footnotetext[1]{Equal contribution}
\renewcommand{\thefootnote}{\arabic{footnote}}
\begin{abstract}
Reusing KV cache is essential for high efficiency of Large Language Model (LLM) inference systems. 
With more LLM users, the KV cache footprint can easily exceed GPU memory capacity, so prior work has proposed to either \emph{evict} KV cache to lower-tier storage devices, or \emph{compress} KV cache so that more KV cache can be fit in the fast memory. 
However, prior work misses an important opportunity: {\em jointly} optimizing the eviction and compression decisions across \emph{all} KV caches to minimize average generation latency without hurting quality. 

We propose \name, a KV-cache management system that applies \emph{lossy compression} and {\em adaptive eviction} to KV cache across multiple storage tiers.
Specifically, for each KV cache of a context, \name considers the effect of compression and eviction of the KV cache on the average generation quality and delay across all contexts as a whole. 
To achieve this, \name proposes a unified \emph{utility function} that quantifies the effect of quality and delay of the lossy compression or eviction. 
To this end, \name's profiling module periodically updates the utility function scores on all possible eviction-compression configurations for all contexts
and places KV caches using a fast heuristic to rearrange KV caches on all storage tiers, with the goal of maximizing the utility function scores on each storage tier. 
Compared to the baselines that evict KV cache or compress KV cache,
\name achieves higher KV-cache hit rates on fast devices, i.e., lower delay, while preserving high generation quality by applying conservative compression to contexts that are sensitive to compression errors. 
Evaluation on 12 datasets and 5 models demonstrates that \name achieves up to 2.19$\times$ faster time-to-first-token (TTFT) at equivalent generation quality.

\end{abstract}

\tightsection{Introduction}

\begin{figure}[ht]
\centering
    \includegraphics[width=\columnwidth]{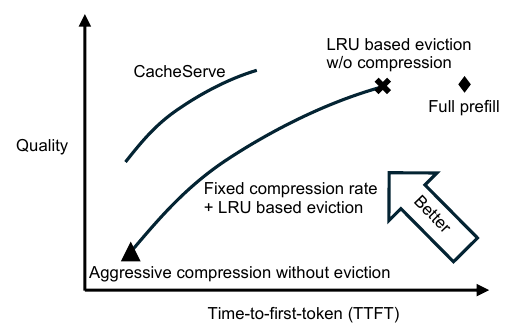}
    \tightcaption{\name jointly optimizes offloading and compression, achieving much better trade-off in terms of quality and TTFT.  This figure is illustrative.   }
    \label{fig:tradeoff}
\end{figure}
Large Language Models (LLMs) are used ubiquitously, in chatbot, agentic, personal assistance, and many other use cases. 
Thus, improving the {\em efficiency of LLM inference}, which provides low-delay experience for users at a low compute cost, is one of the most important problems in both industry and academia. 
To this end, {\em reusing KV cache}---the intermediate states generated during LLM inference---has become the de-facto optimization in modern LLM inference systems. 
By storing and reusing the KV cache of a context in different requests, the inference latency could be significantly reduced, and throughput can be greatly improved~\cite{lmcache, cacheblend, jin2025computeloadkvcache, yang2025kvlinkacceleratinglargelanguage}. 

Realizing the potential performance benefit, however, brings a new system challenge:
the total size of KV cache that needs to be stored continues to grow due to growing model context window limits {\em and} an increased number of concurrent users being served.
Thus, it is challenging to fit all KV cache inside GPU memory. 
Two lines of work exist to avoid re-computing the KV caches that do not fit in GPU memory. 
First, KV cache \emph{eviction}\footnote{KV cache eviction, in this work, means that a KV cache is moved from a faster storage device to the next tier, slower, device. It does not mean deleting the KV cache from the system.} has been proposed to improve the cache hit rate~\cite{h2o, chen2024naclgeneraleffectivekv, qin2025cakecascadingadaptivekv}. 
These works apply classic caching policies, such as least recent used (LRU), to evict KV cache from GPU memory to lower-tier storage devices, which have larger and cheaper space.
This might increase the cache hit rate by allowing more KV caches to be stored and thus avoid repeatedly recomputing the KV cache.
The second line of work applies KV cache \emph{compression} to reduce the memory footprint of each KV cache, allowing more KV caches to fit in the GPU memory~\cite{cachegen, kvquant, kivi, jiang2025kvcomphighperformancellmawarelossy, zhu2025fastcacheoptimizingmultimodalllm}. 

\begin{figure}[t]

\centering
    \includegraphics[width=\columnwidth]{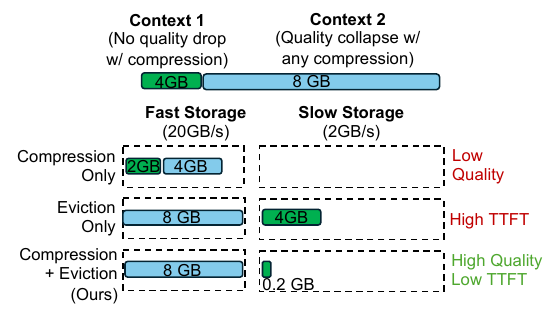}
    \vspace{-6pt}
    \tightcaption{A simple example to illustrate the benefit of jointly deciding compression and eviction. Consider two KV caches (a 4~GB KV cache that can be compressed to 5\% without quality drop, and a 8~GB KV cache where slight compression drops its quality to 50\%), which are originally stored on a fast memory (20GB/s loading bandwidth) and a slow device (2GB/s). A compression-only scheme compresses both KV caches by half, causing low TTFT (0.3 seconds) but low quality (only 75\%); an eviction-only scheme evicts half of the KV cache to slow storage, which has high quality (100\%) but with high TTFT (2.4 seconds) when reusing the evicted KV cache; by considering compression and eviction, we simultaneously achieve high quality (100\%) and low TTFT (0.5 seconds).}
    \label{fig:eviction-compression-illustration}
\end{figure}
However, both eviction and compression approaches are limited when used separately.
When compression is applied, prior works only consider using more aggressive compression when reaching the space limit of the storage devices, and do not consider the possibility of eviction or simply using traditional eviction policies, such as LRU~\cite{cachegen, kvquant, kivi, jiang2025kvcomphighperformancellmawarelossy}.
When eviction is applied, these works do not adapt compression configurations based on the eviction decisions~\cite{h2o, chen2024naclgeneraleffectivekv, qin2025cakecascadingadaptivekv}. 
This is often suboptimal. 
As shown in Figure~\ref{fig:eviction-compression-illustration}, if we consider KV cache for two contexts (context 1 is 4~GB in size and context 2 is 8~GB), context 1 can keep a quality score of 100\% even it is aggressively compressed, while context 2's quality collapses to 50\% even with minimal compression. 
When compression-only is applied, compressing both KV cache to 50\% of their original size, the quality score degrades to 50\% for context 2, with an average quality of 75\% and average TTFT of 0.3 seconds.
When eviction only is applied,  context 1 is evicted to slow device, and context 2 is kept at the fast device, achieving a quality score of 100\% and TTFT of 2.4 seconds. 
A better choice is to combine compression and eviction, specifically keeping context 2 in fast storage, and compressing context 1's KV cache to 0.2GB and evicted to slow storage. This achieves 100\% quality score with TTFT of 0.5 seconds.

Prior works do not consider decisions of eviction or compression on all contexts, primarily due to the huge search space this problem incurs. 
First, when applying compression, compression methods and rates can be adapted. 
Secondly, compression needs to be combined with eviction; for example, after a certain compression rate, eviction should be used to avoid a huge quality drop. 
Furthermore, when dealing with any context's KV cache, considering the impact of compression and eviction on itself is not enough; instead, its impact on other contexts also needs to be considered. 
Compressing a KV cache to a moderate rate and placing it on a fast storage device is beneficial to its quality and TTFT; however, this space can be used to store substantially more KV cache that can be aggressively compressed to improve \emph{overall} TTFT and quality. 

A key insight is that different contexts benefit differently from compression or eviction, since they have various \textit{sensitivities} to lossy compression (\S\ref{subsec:no_compression_for_all}). 
Some contexts are
highly sensitive to compression errors, and even a conservative compression ratio can lead to a huge
quality drop; while others can be largely compressed without affecting generation quality. 
Thus, compression and eviction decisions should be made globally to consider the sensitivities of all contexts in the system.
For example, we should choose to evict highly sensitive contexts that suffer greatly from lossy compression when a higher storage tier is full, while bringing up KV cache from a lower tier that has lower sensitivity to compress them aggressively to fit into the higher storage tier. 



Making global cache management decisions based on all contexts creates a search space to the power of the number of contexts. 
To make it practical, our paper proposes a utility function, taking the storage tier, compression method, and rate as input, outputting a single score that quantitatively measures the effect of each decision on delay and quality. 
This allows us to easily compute the overall delay and quality trade-off based on all contexts for each cache management decision.

Putting it together, we propose \name, a system that adaptively chooses the optimal {\em compression-eviction configuration}---whether to evict, to which storage device, and whether to compress, by what method and at which ratio---on a per KV-cache (per context) basis.
\name first calculates all possible configurations' utility function scores for each context, based on a set of initially generated queries. Then \name puts KV cache that maximizes the total utility function score on a certain storage device. 
Since real-world queries may drift from the initial set of queries, \name records new incoming queries for each context, and performs periodic re-profiling based on the newest set of queries. 
After re-profiling, \name uses a greedy-based algorithm to re-arrange KV cache in different storage tiers, such that the utility function scores can be maximized on each tier.




We implement \name within the existing LLM serving stack by extending vLLM and LMCache with \textasciitilde3K lines of code that add multi-tier KV-cache placement, compression, and eviction control.
Concretely, our implementation intercepts KV-cache lookups, retrievals, and stores, applies per-context compression/eviction policies, and integrates tightly with vLLM’s paged GPU memory management to orchestrate cache movement across GPU, CPU, and SSD tiers.

In short, our contributions are:
\begin{itemize}
    \item We design and implement \name, the first system that jointly considers both lossy compression and eviction for a multi-tier KV cache eviction system.
 \item We propose a {\em utility function} that quantifies the effect of lossy compression and eviction on both quality and delay to make cache management decisions. Using this utility function, \name can compare {\em all} feasible compression-eviction configurations at once with a clear optimization criterion. 
 \item We implement \name in the state-of-the-art inference engine, vLLM~\cite{vllm}, and show that compared to baselines that apply the same compression methods and ratios on all contexts, \name reduces TTFT by 1.43--3.77$\times$ across five different models with the same quality score; compared to the baseline that only applies LRU-based eviction, \name reduces TTFT by 1.22 to 1.56$\times$ within 3\% quality drop, and improves inference throughput by 2.0--3.6$\times$ compared best baseline at a quality score target of 80\%.

\end{itemize}

We should note that the idea of using utility function which captures the effect of delay and quality and adapting compression configurations based on sensitivity are not new~\cite{zhou2025dynamickvtaskawareadaptivekv, yu2025evolkvevolutionarykvcache, kim2025kvzipqueryagnostickvcache, ray2025metis}.
However, we are the first to apply the utility function to enable the joint optimization of eviction and compression, achieving much better results than fixed compression or eviction.  
\tightsection{Background}

\subsection{KV cache reusing and eviction in LLMs}
LLM inference includes two phases, the prefill phase where the entire input context is fed into the LLM at once to produce the \emph{KV cache}, and decode phase where the KV cache is used to generate consecutive output tokens. 
The prefill phase is computation-heavy, especially under long-context scenarios, since its computational complexity grows super-linearly with the input length. 
To accelerate the prefill phase, many recent systems improve inference efficiency by storing and reusing previously computed KV caches to skip redundant computation~\cite{cachegen, lmcache, gu2024llmsteerimprovinglongcontextllm}. Prior works in systems mainly focused on KV cache management inside a hierarchical storage system.

To reduce memory waste, vLLM \cite{vllm} uses PagedAttention to handle the KV cache in smaller, page-like blocks instead of large contiguous chunks, reducing memory fragmentation. SGLang \cite{sglang} optimizes computation and memory usage by employing a tree-based caching mechanism called RadixAttention, which stores KV caches in a radix tree and maintains them via a least recently used (LRU) eviction policy. Some recent works further push the boundary by extending KV cache management across the memory hierarchy and storage. LMCache \cite{lmcache} reuses KV caches for any repeated text, regardless of whether it is a prefix, and offloads KV caches that are not selected to hierarchical caching devices such as CPU memory. Works like Mooncake \cite{qin2025mooncakekvcachecentricdisaggregatedarchitecture}, Attentionstore \cite{gao2024attentionstore}, and InstInfer \cite{pan2024instinferinstorageattentionoffloading} leverage disaggregated resources (i.e., CPU, DRAM, and SSD) to offload KV caches and accelerate LLM serving.


\subsection{KV cache compression}
\label{subsec:compression_background}
Due to the relatively large sizes of KV caches (7.81 GB for 32k token context in Qwen3-32B), much previous work has been done on compressing KV cache.
KV cache compression gives two key benefits in LLM serving. First, by reducing the size of each cache entry, one can store more KV caches within a given memory budget, which increases the probability of reuse and leads to higher cache hit rates. Second, it substantially reduces the data transfer overhead in offloading-based inference systems where KV caches are stored in CPU or SSD memory. This is crucial since transferring large KV caches over bandwidth-limited interconnects can become a performance bottleneck.

There have been many KV cache compression methods, and they differ in both quality retention and system efficiency. These methods can be broadly categorized into: (1) token dropping (e.g., H2O, SnapKV)~\cite{h2o, li2024snapkvllmknowslooking}, which selectively removes less important tokens from the KV cache; (2) quantization (e.g., CacheGen (Adaptive quantization), KIVI (Uniform quantization))~\cite{cachegen, kivi}, which reduces the bit-width representation of KV entries; (3) merging (e.g., HOMER, Look-M)~\cite{song2024hierarchicalcontextmergingbetter, wan2024lookmlookonceoptimizationkv}; (4) prompt compression (e.g., LLMlingua, LongLLMlingua)~\cite{jiang2023llmlingua, jiang2023longllmlingua}. 

Due to the different natures of these categories, these methods have widely different performances. As we demonstrate later in Sec~\ref{sec:compress_dif}, depending on the specific model size, dataset properties, and generation lengths, different methods will have different tradeoffs. Some methods, like CacheGen or KIVI, have a decompression overhead, while others, like H2O, can be directly used in decoding. Moreover, more novel research on KV cache compression is being created on top of previous work every month. Yet there is no principled method to choose and adopt this wide variety of compression algorithms in real systems. 


\tightsection{Motivation}
\label{sec:motivate}
\begin{table}[t]
\centering
\small
\setlength{\tabcolsep}{3pt} 
\begin{tabular}{@{}llll@{}}
\toprule
Dataset     & Length (mean) & Type of contexts & \#Contexts \\ \midrule
Samsum      & 11586         & Few-shot         & 50 \\
TriviaQA    & 12311         & Few-shot         & 50 \\
MultiNews   & 2741          & Multi-document   & 50 \\
2wikimQA     & 7526         & Multi-document   & 50 \\
Qasper      & 4885          & Single-document  & 50 \\
NarrativeQA & 40003         & Single-document  & 50 \\ \bottomrule
\end{tabular}
\tightcaption{Lengths, type and number of contexts used in \S\ref{sec:motivate}. }
\label{tab:dataset}
\end{table}

The most straightforward way of managing KV cache memory is to either apply uniform compression or to use a cache eviction scheme, such as Least Recently Used (LRU) or Least Frequently Used (LFU), on all contexts.
Both compression and eviction can greatly improve inference throughput, either by compressing the memory footprint of KV cache, so it is faster to fetch from remote storage devices~\cite{cachegen, jin2025computeloadkvcache, liu2024minicachekvcachecompression},
or by keeping KV cache that is more likely to be needed in the near future in faster storage devices to make loading faster~\cite{chen2025impress, lmcache, feng2025adaptcachekvcachenative,gao2024attentionstore}. 

However, as we will show in this section, further optimization is possible, as KV cache management decisions should depend on context-specific characteristics. 

In this section, we present an empirical study with six datasets from LongBench dataset~\cite{longbench}, which is a widely-used dataset for verifying LLM's ability to answer long-context questions. 
The dataset details are shown in Table~\ref{tab:dataset}. 
Specifically, we randomly draw 50 contexts from 12 datasets in the LongBench~\cite{longbench} dataset, covering various types of tasks, including few-shot, single-document question answering, and multi-document question answering.

\tightsubsection{Why not compress or evict all contexts? }
\label{subsec:no_compression_for_all}


\begin{figure*}[t]
\centering
\includegraphics[]{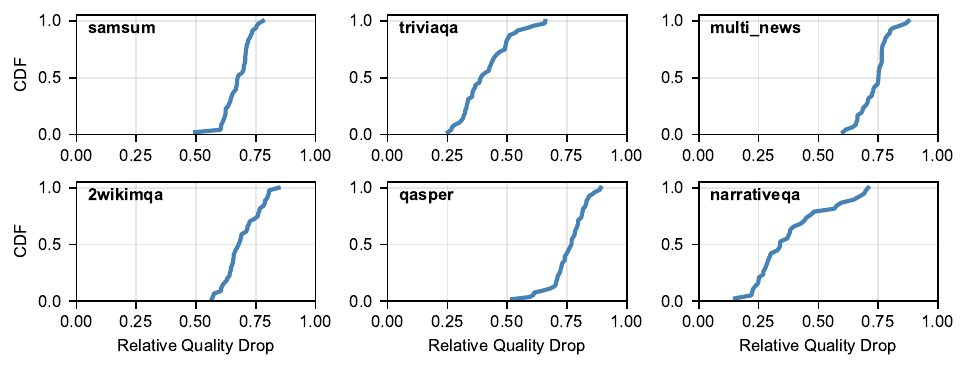}
\tightcaption{Different contexts have different compression error sensitivities in the six datasets mentioned in Table~\ref{tab:dataset}. 
Compression error sensitivity is defined as the quality score drop when applied with keydiff compression at a ratio of 0.9. 
Plotted with Llama-3.1-8B-Instruct. 
}
\label{fig:sensitivity2}
\end{figure*}

KV caches for some contexts are highly sensitive to compression errors: even applying an extremely conservative compression ratio can lead to a substantial quality drop. In contrast, other contexts are minimally affected.
This leads to our first insight as follows. 

\begin{insight}
    Different contexts have different sensitivities to compression errors, thus simply applying the same eviction or compression schemes on all contexts is not optimal.
    \label{insight:adaptive}
\end{insight}

Figure~\ref{fig:sensitivity2} illustrates this, where the CDF of compression error sensitivities, defined as relative quality drop under a compression ratio of 0.9\footnote{Compression ratio is defined as the ratio between compressed KV cache size and original KV cache size. }, is plotted with datasets in Table~\ref{tab:dataset} of different types. 
The figure is plotted with the Llama-3.1-8B-Instruct model with \texttt{keydiff} compression. 
As shown in the figure,
for example, Wikipedia articles (2wikimQA) have high sensitivities, with median of 0.681. 
Our intuition is that critical information is distributed at a high density in all contexts.
Meanwhile, story narratives (NarrativeQA) have a much lower sensitivity, with a median of 0.340, since they are based on dialogues containing redundant information. 

Moreover, we find that even though two datasets belong to the same type of context, they may have completely different distribution of sensitivities. 
For example, Samsum and TriviaQA are both few-shot datasets, where several example question-answer pairs are provided to LLMs before answering the real questions, while Samsum has sensitivity with a median of 0.676 and TriviaQA has sensitivity with a median of 0.392. 
Also, MultiNews samples are multi-document contexts, where multiple news articles are concatenated to form each sample, and Qasper samples are single-document, where each sample only contains a single document on a specific topic. 
However, the MultiNews and Qasper datasets have very similar median sensitivities, 0.738 and 0.759 specifically. 

Overall, sensitivities vary significantly across contexts. 
Specifically, the coefficient of variation (a measure of variance) is from 0.078 to 0.394, and this shows that different contexts have very different sensitivities, even within the same dataset.

\tightsubsection{Why not the same compression method and ratio for all contexts? }
\label{sec:compress_dif}
\begin{figure}[t]
\centering\textbf{}
\includegraphics[width=\columnwidth]{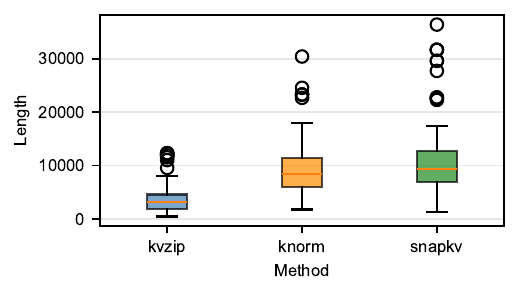}
\tightcaption{Length distributions for contexts whose least quality drop  occurs when using kvzip, knorm, and snapkv, under a compression ratio of 0.6.}
\label{fig:length_compression_method}
\end{figure}

\begin{figure}[t]
\centering
\includegraphics[width=\columnwidth]{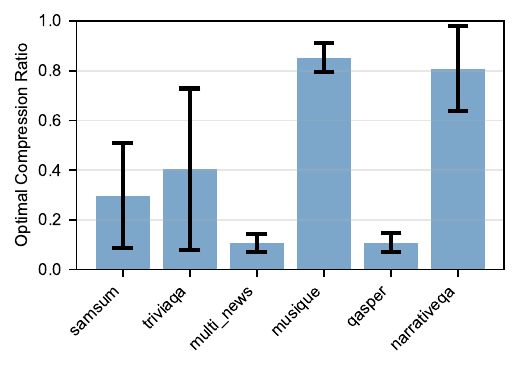}
\tightcaption{Different types of contexts have different optimal compression ratios. Here, each bar shows the mean and standard deviation of optimal compression ratios for different type of contexts. }
\label{fig:compression_ratio_type}
\end{figure}

Another natural question to ask is why not apply a fixed compression method and ratio on all contexts? 
As also discussed in \S\ref{subsec:compression_background}, many parameters can be adjusted, including different compression methods and compression ratios for each method. 
As different methods look at different metrics for dropping information from the KV cache, we hypothesize that different contexts have their own optimal compression method and ratios. 
\begin{insight}
    The optimal compression method and ratio cannot be simply determined by the lengths of contexts or the type of contexts. 
    \label{insight:context_type}
\end{insight}

To verify this, in Figure~\ref{fig:length_compression_method}, we plot the length distribution for three groups of contexts whose optimal compression methods are \texttt{kvzip}, \texttt{knorm} and \texttt{snapkv}, respectively. 
Here, \emph{optimal} compression method is defined as the method with the lowest quality\footnote{Generation quality is defined as the similarity between the generated answer after compression and the original prefill answer.} drop at the same compression ratio. 
There are 40.3\%, 40.0\% and 19.7\% contexts whose optimal methods are \texttt{kvzip}, \texttt{knorm} and \texttt{snapkv}, suggesting that there is no dominant compression method that works for all contexts, and per-context method adaptation is needed. 
Furthermore, we can see that the context lengths distribution for the three groups overlaps with each other, indicating that there is not a clear relationship between context length and which compression method to select. 

Secondly, in Figure~\ref{fig:compression_ratio_type}, we plot the mean optimal compression ratio and the standard deviation for different datasets in Table~\ref{tab:dataset}. 
The {optimal compression ratio} is defined as the ratio with best quality-delay trade-off among all compression methods\footnote{To measure the correlation between TTFT and quality, we use $\text{normalized quality} - \text{normalized TTFT}$ to make sure that quality and TTFT are comparable in terms of scale.}. 
We can see that even within each dataset, the optimal compression ratio varies greatly across different contexts, with coefficient of variation being 0.068 to 0.806. 
This shows that the optimal compression ratios cannot be simply determined by the dataset. 

Moreover, we observe that even though two datasets belong to the same type, such as MultiNews and Musique, which are both multi-document contexts, they have non-overlapping distributions.
This suggests that the optimal compression ratios cannot be trivially determined by type of contexts either.



\section{\name Design}
We now describe the design of \name, starting with an overall workflow (\S\ref{subsec:overall_workflow}), a utility function that determines the quality-rate trade-off of each compression configuration (\S\ref{subsec:util_function}), a profiling mechanism to determine the right compression configuration for each context (\S\ref{subsec:profile}),  and a policy to manage the KV cache when a storage tier is full (\S\ref{subsec:heuristic}).

\tightsubsection{Overall workflow}
\label{subsec:overall_workflow}
\begin{figure}[t]
\centering\textbf{}
\includegraphics[width=\columnwidth]{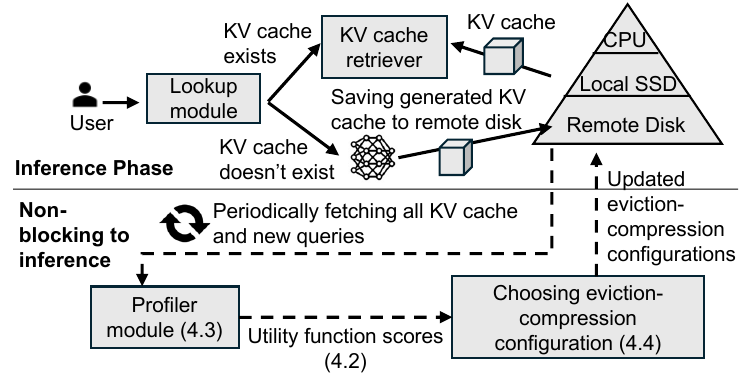}
\tightcaption{End-to-end workflow of \name.}
\label{fig:e2e}
\end{figure}

\name contains two phases: an initial offline phase and an online inference phase.

\mypara{Initial offline phase} During the initial offline phase, \name's profiler (\S\ref{subsec:profile}) computes the utility function scores (\S\ref{subsec:util_function}) for all possible eviction-compression configurations of each context. 
To compute the quality and delay items in the utility function, we use an initial set of generated questions for each context. 

\mypara{Online inference phase} In the online inference phase, when a user request comes in, the lookup module first determines whether the KV cache for the requested context exists in the multi-tier storage system managed by \name. 
If it exists, \name's retriever fetches the KV cache from the location that the lookup module returns. 
If it does not, the KV cache is generated by the LLM, and then saved by \name to the remote disk, which is assumed to have unlimited space. 
As the online queries may differ from the initial set of profile queries, \name's profiling module also performs periodic re-profiling based on an updated set of queries for each context, when the achieved quality is lower than threshold $X\%$ and free GPU cycles are available. 





Note that the KV cache storing operation runs on the CPU, which does not block the online inference computations for other queries. 
Furthermore, the re-profiling step first compresses the KV cache into multiple versions and runs \emph{decoding} phases with different KV cache versions. 
Both the compression step and decoding computations are lightweight in terms of GPU computations, and can be batched with other requests' decoding computations. 
As we will show soon (\S\ref{microbenchmark}), re-profiling incurs minimal system overhead. 

\tightsubsection{Utility function: Balancing quality and loading delay}
\label{subsec:util_function}
Each compression and eviction configuration presents a point on the generation quality and loading delay trade-off. 
To easily measure the impact of each configuration on quality and loading delay, we design a utility function which outputs a single score for each eviction-compression configuration -- the storage tier that the eviction scheme offloads KV cache to, and the compression method and ratio it applies:
\[
\begin{aligned}
\mathrm{Util}(\text{method}, \text{ratio}, \text{device})
    &= (\alpha \cdot \text{quality} - \text{TTFT}) \cdot \text{frequency},
\end{aligned}
\]
where $\alpha$ is the trade-off parameter between quality and loading delay --- higher $\alpha$ improves the quality of \name at the cost of higher loading delay in average, and in our evaluation we will vary $\alpha$ to show the trade-off of \name between quality and loading delay.

In the above equation, the input to the function is the storage device tier which the eviction algorithm sends the KV cache to, and the compression method/ratio that is applied.  The quality is computed based on the compression method and ratio, averaged across all the data in the profiling query set; the delay item is computed with estimated latency to fetch a (compressed) KV cache from the corresponding storage device; and the frequency item is the number of times the context is accessed.

To make it concrete, take the examples from \S\ref{subsec:no_compression_for_all},  when using \texttt{keydiff} compression, across the contexts from 2wikimQA dataset which has higher sensitivity, with compression ratio of 0.6, we can reach a quality score of 0.65 and TTFT of 0.05 seconds; while across contexts from NarrativeQA dataset that has lower sensitivity, with compression ratio of 0.9, we can get a quality score of 0.67 with TTFT of 0.05 seconds. 
Assuming $\alpha=1$ and equal frequency for each context, the utility function score is 0.607 and 0.621 respectively. 
The configuration of (keydiff, 0.6, CPU) is strictly worse than (keydiff, 0.9, CPU) as the former one has the same TTFT and lower quality.
This is reflected by the utility function score too, with the former one having a lower score than the latter. 

The system enables users to navigate the trade-off between generation quality and latency with the tunable parameter $\alpha$. A larger $\alpha$ prioritizes generation quality, while a smaller $\alpha$ prioritizes smaller TTFT. For example, in the 2wikimQA dataset, a compression ratio of 0.4 yields a quality score of 0.82 and a TTFT of 0.20 seconds, whereas a ratio of 0.2 improves quality to 0.92 but increases TTFT to 0.24 seconds. Consequently, if a user sets $\alpha > 0.3$, the utility function favors the higher quality provided by the 0.2 ratio; conversely, a lower $\alpha$ prioritizes the latency reduction offered by the 0.4 ratio.

Another important design question when calculating the utility function is: what is a suitable granularity to compute this utility function on? 
\name computes this utility function based on context-level, instead of chunk-level which splits the whole context's KV cache into multiple chunks, or token-level. 
This is because context-level profiling has a global view of important token distribution considering the entire context. 
On the other hand, for chunk-level profiling, some tokens may be important within the scope of a chunk, while unimportant at the scope of the whole context. 
Token-level profiling is even more suboptimal since quality measurement is not feasible. And token-level retrieval cannot saturate bandwidth and leads to low speed and high overhead because of inconsistent memory access.

\subsection{Profiler module: Computing the utility function scores and choosing eviction-compression configuration}
\label{subsec:profile}

Based on insight~\ref{insight:adaptive}, the compression-eviction configuration must adapt to different contexts. 
Insight~\ref{insight:context_type} indicates that the configuration cannot simply be determined by the characteristics of the context, such as lengths or types. 
Thus, instead of coming up with context characteristics that indicate the benefits of each configuration, \name directly profiles the impact of all configurations on overall \emph{quality} and \emph{delay}. 
Specifically, \name uses the utility function (\S\ref{subsec:util_function}) that outputs a single score for each configuration of (storage device, compression method, compression ratio) to capture both delay and quality.

Before the online inference stage, 
as the system does not see any queries before, the profiling module generates a set of training questions for each context following prior works~\cite{catridges, wei2024simplesyntheticdatareduces, zhang2024inftybenchextendinglongcontext, wei2024longformfactualitylargelanguage, white2025livebenchchallengingcontaminationlimitedllm}, to compute the average quality and delay on them. 
The questions are generated by the GPT-5 API about different paragraphs in the contexts, aiming to obtain the eviction-compression configuration that drops as much redundancy for each paragraph as possible to reduce delay. 


During online inference, as real-world queries may drift from the training queries, the profiling module also periodically re-profiles with a newer dataset. 
The re-profiling is triggered whenever the quality achieved on the online testing dataset is below the quality achieved on the profiling dataset by a threshold of $X\%$, and when the system has free GPU cycles.
Note that the profiling process only requires the KV cache to be compressed with different methods and ratios, which incurs very minimal overhead, and running the decoding phase with these different compressed versions. 
The decoding phase is known to be largely batchable (batching with other requests does not increase decoding delay).  
Thus, available GPU cycles occur when the number of online inference requests is smaller than the maximum batch size in the system.  
In practice, users can freely adapt $X\%$ based on their quality requirements.

\subsection{Eviction-compression configuration selection module}
\label{subsec:heuristic}

Given an initial set of contexts, where each context has a set of generated queries for profiling, \name computes the utility function scores for all possible eviction-compression configurations on each context. 
\name's configuration selection module then places the (compressed) KV cache to a certain storage tier for each context with the configuration corresponding to the highest utility function score. 

\begin{figure*}[t]
\centering
    \includegraphics[]{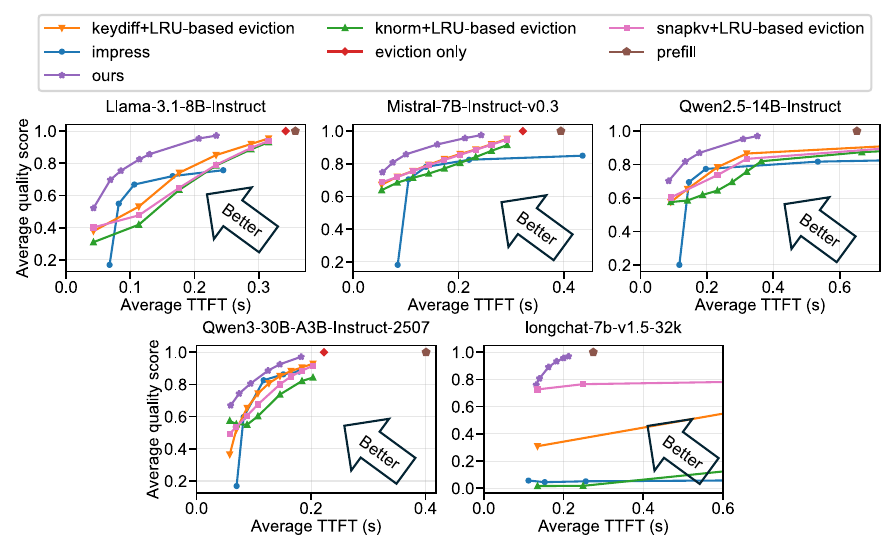}
    \tightcaption{\name reduces TTFT by 1.43 to 3.77$\times$ with the same quality score compared to compression + LRU-based eviction, and reduces TTFT by 1.22 to 1.56$\times$ compared to prefill and eviction only scheme. }
    \label{fig:score_vs_ttft1}
\end{figure*}

When a storage device $S$ is full, existing KV cache on that device needs to be updated.
Specifically, each context's KV cache can be updated either by more aggressive compression to fit in $S$, or by evicting it to a lower-tier storage to save space. 
Each context has $(\#\text{compression methods}\times\#\text{compression ratios}\times\#\text{lower-tier storage devices})$ of options to choose from. 
Here, we filter out options that keep KV cache on storage $S$ but do not save space (such as keeping the compression ratios and not evicting to the next tier).


The problem of deciding the configuration for each context while maximizing the total utility function score on device $S$ is a Multi-Choice Knapsack Problem, which is NP-hard. 
We greedily solve this problem by repeating the greedy process of finding updated configurations with the lowest utility score drop for contexts on $S$, until the updated KV cache fits successfully. 
Then, the same update process is applied recursively if the lower-tier storage is also full.
The exact greedy algorithm for this process is described in \S\ref{pseudo}.

\section{Implementation}

We implement \name on top of vLLM \texttt{v0.11.2} and LMCache \texttt{v0.3.9post2}, with about 3K lines of code in Python based on PyTorch \texttt{v2.9}.

When a new text query comes in, we call LMCache's KV cache lookup module, specifically \texttt{lookup(tokens)}.
If it finds that the requested KV cache exists in the system, the number of hit tokens is returned, and the vLLM engine allocates the amount of GPU memory for storing the KV cache for the hit tokens. 
Then, LMCache's \texttt{retrieve(tokens)-->KV Cache} function is triggered to load the KV cache from \name-managed multi-tier storage backend, and put it into the paged GPU memory allocated by vLLM. 

If it does not find the requested KV cache, LMCache's \texttt{store(tokens, KV Cache)} function is triggered to save the generated KV cache to CPU DRAM, and then \name calls the \texttt{manage} function, which places the new KV cache on a storage device and updates existing KV cache entries if the device is full. 
Specifically, we use a class \texttt{KVConfig(KV Cache)} which records the utility function scores for all possible configurations for the given KV cache. 
The \texttt{manage} function takes in \texttt{[KVConfig(KV Cache 1), ..., KVConfig(KV Cache n)]} and computes \texttt{(device placement, compression method, compression ratio)} for each KV cache that needs updating, calculated by \name's configuration selection module (\S\ref{subsec:heuristic}). 
Finally, the \texttt{manage} function updates those KV caches with the new configuration. 



\begin{figure*}[t]
\centering
    \includegraphics[]{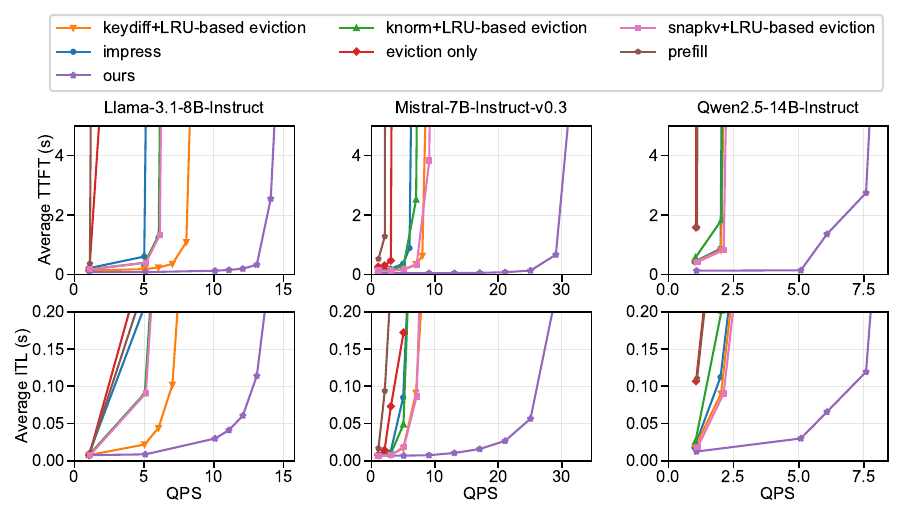}
    \tightcaption{\name maintains consistently low TTFT and ITL across all QPS levels.}
    \label{fig:qps_vs_ttft}
\end{figure*}

\section{Evaluation}
The key takeaways from the evaluation are: 

\begin{itemize}
    \item Across 5 models and 555 contexts from 12 datasets from LongBench~\cite{longbench}, \name reduces TTFT by 1.43 to 3.77$\times$ and improves quality by 13.58\% to 55.40\% at the same TTFT, compared to the baseline that combines fixed compression and classic eviction. 
    \item Compared to smart KV cache eviction system IMPRESS\cite{chen2025impress}, \name reduces TTFT by 1.5 to 5.2 $\times$ and improves quality by 14.29\% to 27.00\% at the same TTFT. 
    \item Compared to full prefill or full eviction, \name reduces TTFT by 1.29 to 2.19$\times$ within 3\% of quality drop.
    
\end{itemize}

\tightsubsection{Evaluation setup}

\begin{table}[htbp]
\centering
\footnotesize
\begin{tabular}{lrr p{4cm}}
\hline
\textbf{Dataset} & \textbf{Mean} & \textbf{Std} & \textbf{Topic} \\ 
\hline
narrativeqa & 108K & 55K & Long stories (literature, film).  \\
qasper & 24K & 12K &  NLP research papers \\
multifieldqa\_en & 29K & 15K & Legal documents and reports.  \\
hotpotqa & 57K & 18K & Wikipedia articles. \\
2wikimqa & 30K & 15K &  Wikipedia articles.  \\
musique & 69K & 9K &  Wikipedia articles.  \\
gov\_report & 54K & 34K & US GAO and CRS reports.  \\
qmsum & 57K & 27K & Meeting scripts and records.  \\
multi\_news & 12K & 10K & News articles on the same event. \\
trec & 30K & 12K & News articles and web pages.  \\
triviaqa & 47K & 25K & Trivia questions.  \\
samsum & 34K & 17K & Messenger-like conversation histories.  \\ 
\hline
\end{tabular}
\caption{Dataset Statistics and description. }
\label{tab:datasets_eval}
\end{table}

\mypara{LLMs}
We cover a wide range of popular open-source LLMs, including dense LLMs, including Llama (meta-llama/Llama-3.1-8B-Instruct), Qwen (Qwen/Qwen2.5-14B-Instruct), LongChat (lmsys/longchat-7b-v1.5-32k) and Mistral (mistralai/Mistral-7B-Instruct-v0.3), and also Mixture-of-Expert LLM (Qwen/Qwen3-30B-A3B-Instruct-2507). 

\mypara{Dataset}
We use LongBench benchmark~\cite{longbench}, which combines many long-context datasets and includes a wide range of context genres (including government reports, news, conversations and stories), as our evaluation dataset. 
The dataset statistics, including the mean and standard deviation of the context lengths, and the dataset description is listed in Table~\ref{tab:datasets_eval}. 

\mypara{Query construction}
We randomly sampled 555 contexts from longbench dataset.
For each context, we prompts gpt-5 to generate 100 QA queries (following previous works~\cite{catridges, wei2024simplesyntheticdatareduces, zhang2024inftybenchextendinglongcontext, wei2024longformfactualitylargelanguage, white2025livebenchchallengingcontaminationlimitedllm}), 50 as training queries to build the profiler of \name and 50 for testing. 

\mypara{Hardware}
We run our experiments on one 80GB H100 GPU that hosts the model weights for each model. 
We allocate 80GB CPU DRAM, 800GB SSD, and assume that the remote storage has unlimited space. 



\mypara{Baselines} 
We compare \name with baselines that apply KV cache compression or eviction.
\begin{packeditemize}

\item \textit{Prefill}: This baseline performs full prefill on each query without prefix caching. We run this with vLLM \texttt{v0.11.2}. 

\item \textit{Eviction only}: When a storage tier becomes full, this system evicts KV cache to the next storage tier using an least-recently used (LRU) policy, without any compression.

\item \textit{(keydiff/knorm/snapkv) Compression + LRU-based eviction}: This baseline applies either keydiff~\cite{park2025keydiffkeysimilaritybasedkv}, knorm~\cite{devoto2024simpleeffectivel2normbased} or snapkv~\cite{li2024snapkvllmknowslooking} compression on all contexts. 
We also vary the compression ratio for each method, which forms the quality-TTFT trade-off. 
At a certain compression ratio, when a storage tier is full, the system evicts the compressed KV cache to lower-tier storage with LRU. 
Among these methods, keydiff computes the average cosine similarity of every token to all other tokens, and drops the tokens whose similarities are high among all; 
knorm computes the L2 norm of the key cache, and drops those with low L2 norm; 
snapkv computes the cross attention between tokens near the query and the earlier contexts, and drops those tokens with low cross-attention scores.  

\item \textit{IMPRESS}:
IMPRESS~\cite{chen2025impress} uses three out of eight attention heads to compute the attention scores to profile important tokens in the KV cache. 
IMPRESS keeps $X\%$ of important tokens whose attention scores are among the top $X\%$ of all tokens. 
Then it splits each KV cache into chunks, and only loads KV cache chunks containing important tokens during inference. 
We vary $X\%$ in our experiments to achieve the quality-delay trade-off of IMPRESS. 


\end{packeditemize}

\mypara{Evaluation metrics} 
\begin{packeditemize}
        \item \textit{Quality score}: measures the similarity of answers based on compressed KV cache vs. answers based on uncompressed KV cache. Following prior work~\cite{arabzadeh2024adaptingstandardretrievalbenchmarks, gaikwad2024generative, reimers-2019-sentence-bert}, we compute the cosine similarity between the embedding of the generated answer and that of the answer based on uncompressed KV cache using the \textit{MiniLM-L6-v2} model.
    The higher this score is, the closer the generated answers is compared to answers on uncompressed KV cache, which is the better. 
    
    \item \textit{Time-to-first-token (TTFT)} is the time between the arrival of a query to the time of the first generated token.
    This includes the delay to retrieve KV cache from the storage devices, and the time to prefill the new queries.
    
    \item \textit{Inter-token-latency (ITL)} is the average delay for generating consecutive tokens. 
    ITL reflects the encoding throughput, and increases when the inference engine has more running requests.

    \item \textit{End-to-end latency} is the time between when the request comes in and when the last token is received.
    

\end{packeditemize}

\begin{figure}[t]
\centering
    \includegraphics[width=\columnwidth]{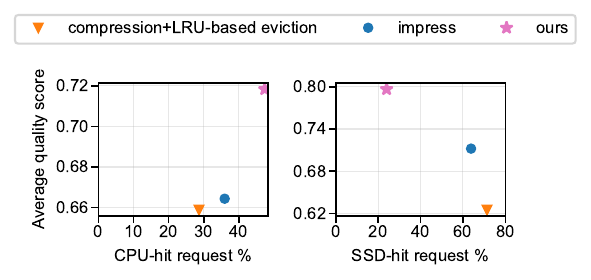}
    \tightcaption{With \textit{Qwen2.5-14B-Instruct}, \name has higher fast device (CPU) hit request percentage while maintaining higher quality.}
    \label{fig:why_we_are_good}
\end{figure}

\subsection{Evaluation results} 

\mypara{TTFT v.s. generation quality}
In Figure \ref{fig:score_vs_ttft1}, compared to 

\noindent the baselines that applies keydiff, knorm and snapkv and 

\noindent LRU-based eviction, \name reduces TTFT by 1.43 to 3.77$\times$ at the same quality score, and improves quality scores by 13.58\% to 55.40\% at the same TTFT. 
Compared to full prefill that does not affect generation quality, \name reduces TTFT by 1.29 to 2.19$\times$ within 3\% of quality drop.
Compared to LRU-based eviction, \name reduces TTFT by 1.22 to 1.56$\times$ with quality drop less than 3\%. 
Finally, compared to IMPRESS, \name reduces TTFT by 1.5 to 5.2$\times$ without degrading generation quality, or improves generation quality by 14.29\% to 27.00\% without increasing TTFT. 

\mypara{TTFT and ITL v.s. QPS}
Figure~\ref{fig:qps_vs_ttft} shows the TTFT and ITL under different query-per-second (QPS), at a quality score target of 80\% for Llama-3.1-8B-Instruct, Mistral-7B-Instruct-v0.3 and Qwen2.5-14B-Instruct. 
At the same TTFT level, \name achieves 2.0 to 3.6$\times$ higher request processing rate, compared to the best baseline. 
Similarly, at the same ITL level, \name achieves 1.9 to 3.4$\times$ higher QPS compared to the best baseline.

\mypara{\name's improvement} \name outperforms different baselines for different reasons. 
First, compared to the baselines that apply same compression methods and ratios on all contexts, \name is better as the context sensitivity difference is considered and eviction-compression configuration is adjusted for different contexts. 
Secondly, compared to LRU-based eviction, which evicts all contexts that cannot fit on the CPU DRAM to lower tier storage, \name is better since some low-sensitivity contexts can be largely compressed without being evicted to slow storage. 
Thirdly, compared to full prefill which incurs significant amount of GPU computation, \name is better since the prefill computation is  skipped. 
Finally, compared with IMPRESS, which splits KV cache into chunks, and not only loads important tokens, but also unimportant tokens in each chunk to fast device, \name only puts KV cache that maximize the total utility function scores on fast device. 

In summary, \name achieves much higher cache hit rate in CPU DRAM, as shown in Figure~\ref{fig:why_we_are_good}, compared to baselines that apply the same compression methods and LRU-based eviction on all contexts and IMPRESS. 
On the other hand, as shown in the right hand side of the figure, the cache hit rate on local SSD for \name is lower than other baselines, which illustrates that most of the cache hit tokens reside on CPU DRAM, providing much faster loading delay than baselines. 

We acknowledge that the improvement of \name depends on the sizes of KV cache for different LLMs. 
When the KV cache is small, DRAM can hold most tokens, leaving little room for improvement. For example, the MoE model Qwen3-30B-A3B-Instruct-2507 has a small KV cache (0.0915 \,GB per 1K tokens), while a smaller dense model such as Llama-3.1-8B-Instruct has 0.12GB KV cache per 1K tokens, so even with a conservative compression ratio, most of the KV cache can be stored in CPU DRAM, leading to low TTFT.

\subsection{Microbenchmarks}\label{microbenchmark}

\begin{figure}[t]
\centering
\includegraphics[width=\columnwidth]{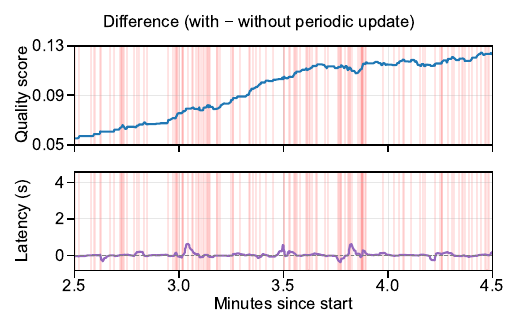}
\tightcaption{Quality and latency over time under Azure workload\cite{wang2025burstgptrealworldworkloaddataset, shahrad2020serverless, zhang2021faster} with only one training query per context. \textbf{Red vertical lines} denote periodic profiling events with \textit{Llama-3.1-8B-Instruct}. Frequent profiling at the beginning and around 3.75–4 minutes introduces brief latency spikes, but latency quickly returns to baseline. As profiling corrects stale compression decisions, model quality steadily improves throughout execution.}
\label{fig:period}
\end{figure}

\mypara{The effect of periodic update} Figure \ref{fig:period} evaluates our periodic-profiling mechanism (\S\ref{subsec:profile}) using real timestamps from the Azure inference trace~\cite{wang2025burstgptrealworldworkloaddataset}.
As Azure inference trace does not include the concrete content of the request, we use the request we generated (described in \S\ref{subsec:profile}) instead.
\name then performs profiling when the discrepancy between the profiled quality and the predicted quality of the query decoded with the stored KV cache exceeds 0.3.

We illustrate the time series of using periodic profiling versus not in Figure \ref{fig:period}:
\begin{packeditemize}
  \item The figure above shows how the quality gain of using periodic profiling versus not evolves over time.
  From this figure, we can see that the quality gain increases gradually over time and eventually saturates at around 11\%, because periodic profiling allows \name to use up-to-date profile to make near-optimal compression decision, while the profile of not doing periodic update gradually gets outdated and result in lower accuracy under the same storage resource budget.
  \item The figure below shows how the end-to-end latency of requests of \name vaaries over time. We can see that, \name has higher end-to-end latency in the periodic profiling intervals (the red intervals) in average, because periodic profiling requires GPU resources and slow down the execution of existing requests.
  That said, the quality improvement brought by periodic profiling (about 10\%) still justifies its cost.
\end{packeditemize}

\begin{figure}[t]
\centering
    \includegraphics[width=\columnwidth]{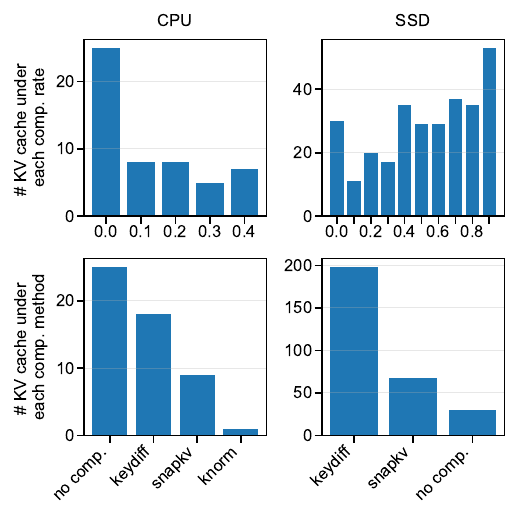}
    \tightcaption{Distribution of selected compression methods and compression rates across CPU and SSD storage tiers under $\alpha=1$ with \textit{Mistral-7B-Instruct-v0.3}. \name actively uses a broad combination of configurations, validating the need for a large search space.}

    \label{fig:search_space}
\end{figure}

\mypara{Different context needs different eviction-compression configurations}
Figure~\ref{fig:search_space} visualizes the distribution of selected eviction-compression configurations. 
We observe that the system indeed makes use of a diverse set of compression configurations, indicating that a large search space is not only necessary but actively exploited. 
We highlight two insights:
\begin{packeditemize}
    \item KV caches on disk actually requires higher compression ratio in average compared to KV caches on CPU.
    This is because KV caches on disk are slower to read, which requires \name to further compress the KV caches to reduce their loading time.
    \item We note that, though \textit{keydiff} generally outperforms \textit{snapkv} (which in turn outperforms \textit{knorm}), yet a non-negligible portion of KV caches still adopts \textit{snapkv} due to context-specific benefits. 
\end{packeditemize}

\begin{figure*}[t]
\centering
    \includegraphics[]{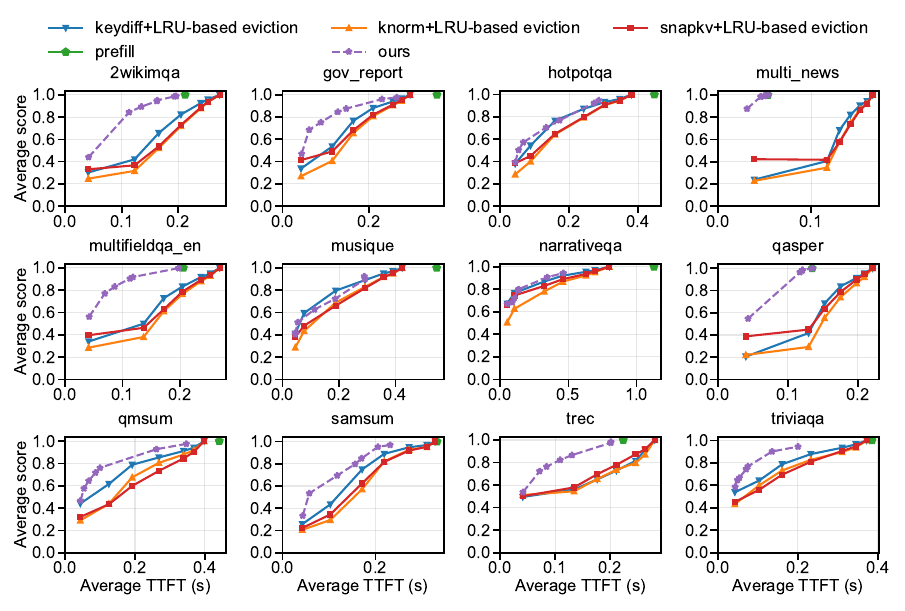}
    \tightcaption{\name's consistently improves TTFT-quality trade-off on different datasets. 
    Among these datasets, \name has better improvement on shorter contexts (such as \texttt{multi\_news} and \texttt{qasper}). }
    \label{fig:qps_vs_ttft_12}
\end{figure*}



\mypara{Improvements across heterogeneous contexts}
Figure \ref{fig:qps_vs_ttft_12} illustrates the TTFT --- Average score trade-off of \name and baselines across different subtasks of longbench dataset.
We draw two conclusions:
\begin{packeditemize}
    \item \name has higher improvement on tasks that corresponds to shorter contexts (\eg \texttt{multi\_news} and \texttt{qasper}).
    This is because, in order to achieve higher CPU cache hit rate, \name puts shorter contexts in CPU instead of longer ones, so that the CPU can fit into the KV caches of more requests.
    \item \name improves the performance for 11 out of 12 tasks. The only exception is on \texttt{musique}, where \name performs slightly worse than the keydiff with LRU eviction baseline.
\end{packeditemize}



\tightsection{Related Work}
\mypara{KV cache reuse and eviction systems} Storing and reusing KV cache across multiple requests has been a critical focus in the design of high-performance LLM serving systems. 
This includes paged memory management in GPU, such as vLLM~\cite{vllm} and SGLang~\cite{sglang}. 
However, this line of work suffers from low cache hit rate when the request arrival rate is high and the number of concurrent users is large, since GPU memory only has limited space (40 to 80~GB for A100 or H100 GPUs). 
Thus, more work is proposed to evict KV cache that cannot fit in GPU memory to lower-tier storage devices, such as CPU DRAM~\cite{lmcache, xiong2024layerkvoptimizinglargelanguage, sglang, li2025continuumefficientrobustmultiturn}, local disks~\cite{zhang2025kvswapdiskawarekvcache, jin2025computeloadkvcache, lmcache}, and even remote disks~\cite{jin2025computeloadkvcache, lmcache}. 
For example, LMCache \cite{lmcache,jiang2025networking} reuses splits each KV cache into multiple chunks, and optimizes the loading speed from CPU/disks to GPU memory with high-performance CUDA kernels; Mooncake \cite{qin2025mooncakekvcachecentricdisaggregatedarchitecture} and InstInfer \cite{pan2024instinferinstorageattentionoffloading} optimizes the loading bandwidths across different GPU clusters and performance of prefill-decode disaggregation; IMPRESS applies fine-grained token-level eviction, which uses a subset of attention heads to determine important tokens from the KV cache, and evicts non-important ones to lower-tier storage~\cite{chen2025impress, xie2025strata}. 
However, this line of work only applies different eviction schemes on KV cache, not even considering combining with compression schemes to achieve higher cache hit rates. 

\mypara{KV cache compression} To achieve higher cache hit rate, many works have explored orthogonal methods to compress KV caches to reduce memory footprint. 
Token dropping methods drop less important tokens from the KV cache, either based on attention scores or learned importance signals~\cite{h2o, chen2024naclgeneraleffectivekv, qin2025cakecascadingadaptivekv}. Quantization methods reduce the size of KV cache by using fewer bits to represent each KV cache element, including adaptive quantization where different layers are applied with different levels of quantization~\cite{cachegen, kvquant} and uniform quantization~\cite{kivi}. 
Similarly, this line of work only considers the possibility of compression, without combining it with eviction, even when the quality drop is high. 
Another line of work applies different levels of compression considering the error sensitivity of each context~\cite{cheng2025qaq, xiong2025uncompmatrixentropyuncover}. 
However, it does not consider per-context sensitivity difference to the problem of joint optimization between eviction and compression as \name does. 

\mypara{Faster LLM serving} Another complementary line of work speeds up LLM serving by better scheduling to reduce the bubbles across different inference stages (\eg prefill, decode or tool-calling)~\cite{agrawal2023sarathi,cui2025optimizingsloorientedllmserving}; prefill-decode disaggregation to place the prefill and decode phases to different GPUs to reduce SLO violation rate~\cite{zhong2024distserve, patel2024splitwiseefficientgenerativellm}; 
speculative decoding approaches which proposes using a smaller, faster draft model to propose a sequence of future tokens, and using the larger target model to verify the tokens in parallel~\cite{Leviathan2022FastIF, miao2023specinfer};
and chunked-prefill to enable piggy-backed decoding and reduce memory consumption of prefill~\cite{agrawal2023sarathi}. 
This line of work is complementary and orthogonal to \name, and can be combined with \name to further increase the efficiency of LLM inference.

\mypara{Other KV cache related optimizations} There are a lot of other works that optimize KV cache in LLM inference, including non-prefix KV cache sharing with approximation~\cite{cacheblend, gim2023prompt}, cross-LLM KV cache reusing with partial layer recomputation~\cite{liu2025droidspeakkvcachesharing}, request routing that is aware of KV cache location~\cite{srivatsa2024prebleefficientdistributedprompt}, improving paged attention for multi-modality models with different shapes of KV cache~\cite{tu2024vlcachesparsitymodalityawarekv}, and managing KV cache for sparse-attention models~\cite{cao2025sparseattentionmultiplecontextkv}.
While this line of work also makes KV cache in LLM inference more practical in multiple use cases, \name is complementary to them as they jointly optimize eviction and compression. 

\tightsection{Limitations and future work}

\mypara{Deployment scope and assumptions}
\name targets a common and challenging deployment scenario with limited GPU and CPU memory and a multi-tier storage hierarchy, where KV caches routinely contend for capacity, and eviction or compression decisions concretely affect end-to-end latency and quality. 
In settings with abundant DRAM (\eg models with small KV cache size or workloads where all KV caches can be kept in fast memory), the advantage of sophisticated joint optimization over simpler policies naturally diminishes. 
Moreover, when the loading bandwidths for fast and slow storage devices are similar, \name's benefits of joint optimization will decrease. 
Our prototype is evaluated on a single-node GPU setup. 
Extending \name to multi-node deployments, cross-GPU cache sharing, highly multi-tenant environments with strict isolation requirements, and alternative serving systems and storage hierarchies is an important direction for future work.

\mypara{Adapting to changing loading bandwidth}
\name currently assumes that the bandwidth of each storage tier is known and relatively stable over the time at which we profile and re-profile configurations. 
When the available bandwidth changes more rapidly than our re-profiling frequency or when bandwidth fluctuations are strongly correlated with bursty query arrivals, the precomputed utility function scores could temporarily deviate from the true quality-latency trade-offs and lead to suboptimal decisions.
As part of our future work, we plan to extend \name to react more quickly to bandwidth shifts, incorporate richer online telemetry, and co-design bandwidth-aware profiling and scheduling policies.

\mypara{Extending to a richer set of compression methods} In the evaluation section, \name selects compression methods from three methods that drop important tokens from KV cache to shrink its memory footprint. 
\name can be easily extended to choose from KV cache quantization methods~\cite{kivi, kvquant, he2024zipcacheaccurateefficientkv}.
Similarly, we can build a profiling configuration pool that combines both token dropping and quantization, and adapts the fraction of tokens to be dropped and the number of bits used to represent each KV element.



\tightsection{Conclusion}
We present CacheServe, a caching system for large language model inference workload that co-designs compression and eviction policies under  hierarchical storage devices to achieve better quality-delay tradeoff. 
CacheServe's core innovation lies in its capability to jointly exploit the benefit of both eviction and compression, by defining a utility function and selecting the compression-eviction that maximizes such utility.
We implement our system on top of production level LLM serving engine (vLLM)  and cache engine (LMCache).
Our evaluations across diverse models and datasets demonstrate that CacheServe significantly reduces average Time-To-First-Token by 1.43 to 3.77$\times$ with equivalent quality as baselines or improves quality by 13.58\% to 55.40\% at the same TTFT. By designing our forward-compatible architecture for future KV cache compression and eviction techniques, \name opens up a new dimension for LLM inference optimizations.

\bibliographystyle{plain}
\bibliography{citations} 

\pagebreak
\appendix
\section{Pseudo Code for Algorithm}\label{pseudo}
\begin{algorithm}[h]
    \caption{Policy Optimizer}
    \label{alg:example}
    \begin{algorithmic}[1]
        \REQUIRE Incoming KV Cache that needs to be stored $i$ (size of the kv cache $l_i$, quality function $Q_i(compression \ ratio)$), Storage queue $T$, Current Occupied size $S_T$, Storage memory size $C$\\
        \ENSURE List of operations on KV Cache: [( KV Cache $i$, compression ratio $r_i$, operation $o_i$)] 
        \STATE $r_i \gets \arg\max Q_i(r)$
        \STATE $T$.append$(i, r_i)$
        \STATE $L$.append$((i, r_i, insert))$
        \STATE $S_T = S_T +sizeof (l_i * r_i)$
        \WHILE{$S_T>C$} 

            \STATE $j, r, o, \gets$ KV Cache in $T$ with the least utility drop operation and the operation. 
            \IF{$o == compress$}
                \STATE $S = S_T - T_j * (r_j - r)$
            \ELSIF{$o == evict$}
                \STATE $r \gets 0$ 
                \STATE $S = S_T - (T_j * r_j)$
            \ENDIF
            \STATE $L$.append$((j, r, o))$
        \ENDWHILE 
        \RETURN $L$ to executor.        
    \end{algorithmic}
\end{algorithm}
\end{document}